\documentstyle[epsfig,preprint,aps,pra]{revtex}
\tightenlines

\begin{document}


\title{Inelastic Collision Rates of Trapped Metastable Hydrogen}

\author{David~Landhuis,\cite{DLaddr} Lia~Matos, Stephen~C.~Moss,
Julia~K.~Steinberger, Kendra~Vant, Lorenz~Willmann,\cite{LWaddr}
Thomas~J.~Greytak, and Daniel~Kleppner}

\address{Department of Physics and Center for Ultracold Atoms, \\
Massachusetts Institute of Technology, Cambridge, Massachusetts 02139}

\date{\today}

\maketitle
 
\begin{abstract}

We report the first detailed decay studies of trapped metastable
($2S$) hydrogen.  By two-photon excitation of ultracold H samples, we
have produced clouds of at least $5\times10^7$ magnetically trapped
$2S$ atoms at densities greater than $4\times10^{10}$~cm$^{-3}$ and
temperatures below 100~$\mu$K.  At these densities and temperatures,
two-body inelastic collisions of metastables are evident.
Experimental values for the total two-body loss rate constant are
$K_2=1.8^{+1.8}_{-0.7}\times10^{-9}$~cm$^3$/s at 87~$\mu$K and
$K_2=1.0^{+0.9}_{-0.5}\times10^{-9}$~cm$^3$/s at 230~$\mu$K.  These
results are in the range of recent theoretical calculations for the
total $2S$-$2S$ inelastic rate constant.  The metastable clouds were
excited in a gas of ground state ($1S$) hydrogen with peak densities
reaching $7\times10^{13}$~cm$^{-3}$.  From the one-body component of the
metastable decay, we derive experimental upper limits for $K_{12}$, the rate
constant for loss due to inelastic $1S$-$2S$ collisions.

\end{abstract}



\section{Introduction}


The ability to create large clouds of metastable ($2S$) H atoms in a
magnetic trap makes it possible to study the rich collisional physics
of cold metastable H.  In an encounter between two metastables,
the possible outcomes include Penning ionization, the formation of
molecular ions, excitation transfer to short-lived $2P$ states, and
hyperfine transitions \cite{fcd00,skv88}.  If the $2S$ atoms are
generated from a background $1S$ gas, then several $1S$-$2S$ collision
processes may also occur.  Though these states of H are among the
simplest atomic states in nature, an accurate description of their
inelastic collisions at low energies remains a theoretical challenge
\cite{fcd00,jsp02,jdd96}.

At low temperatures, the rate of quenching collisions between
metastables is small enough to allow dense ($>10^{10}$~cm$^{-3}$) $2S$
clouds to exist for tens of milliseconds, which is an appreciable
fraction of the 122~ms natural lifetime of the $2S$ state.  Such a
cloud can serve as an atom source for precision spectroscopy of
transitions originating in the $2S$ state.  In particular, the absolute
frequencies of transitions from the $2S$ to higher-lying states can be
combined with the well-known $1S$-$2S$ interval to simultaneously
determine the $1S$ Lamb shift (a sensitive test of QED) and the
Rydberg constant (which relates several more fundamental constants)
\cite{uhg97,sjd99,nhr00}.  Taking advantage of the high
signal-to-noise ratio possible in a cold, trapped sample, spectroscopy
of metastable H may lead to a significant advance in accuracy for the
determination of these quantities \cite{wk00}.

Cold metastable H is also potentially interesting for ``quantum atom optics''
experiments involving single-atom detection.  The large (10~eV)
internal energy of a $2S$ atom is easily registered by an ionizing
collision on an electron-multiplying detector, or by detecting the
Lyman-$\alpha$ photon emitted at a surface or other localized electric
field.  As demonstrated already with metastable noble gases,
single-atom detection allows experiments involving atom-atom
correlations, atom interferometry, or atom holography
\cite{ys96,myk96,rsb01}.  A metastable beam ejected from a trapped H
sample, potentially much brighter than other cold metastable beams to
date, may be a useful source for quantum atom optics research.


After improving our apparatus at MIT, we have been able to generate
metastable H clouds which are larger and longer-lived than in our
previous work \cite{cfk96,kfw98}.  In this paper, we report results
from decay studies of trapped $2S$ clouds, conducted at temperatures
ranging from a few mK to below 100~$\mu$K.  These results include the
first experimental determinations of the $2S$-$2S$ two-body loss rate
constant $K_2$ and upper limits for the $1S$-$2S$ inelastic rate
constant $K_{12}$.  In the case of $K_2$, we have observed a
temperature dependence below 230~$\mu$K which is not predicted in the
current theory for cold metastable H collisions \cite{fjs02}.  This
work serves as a stepping stone to high-resolution spectroscopy
experiments which probe fundamental physics.  Knowledge of metastable
collision parameters may also be instrumental in the development of
single-atom detection experiments involving H.

\section{Experimental Methods}
\label{sec:methods}

\begin{figure}
\vspace{0.5in}
\centering\epsfig{file=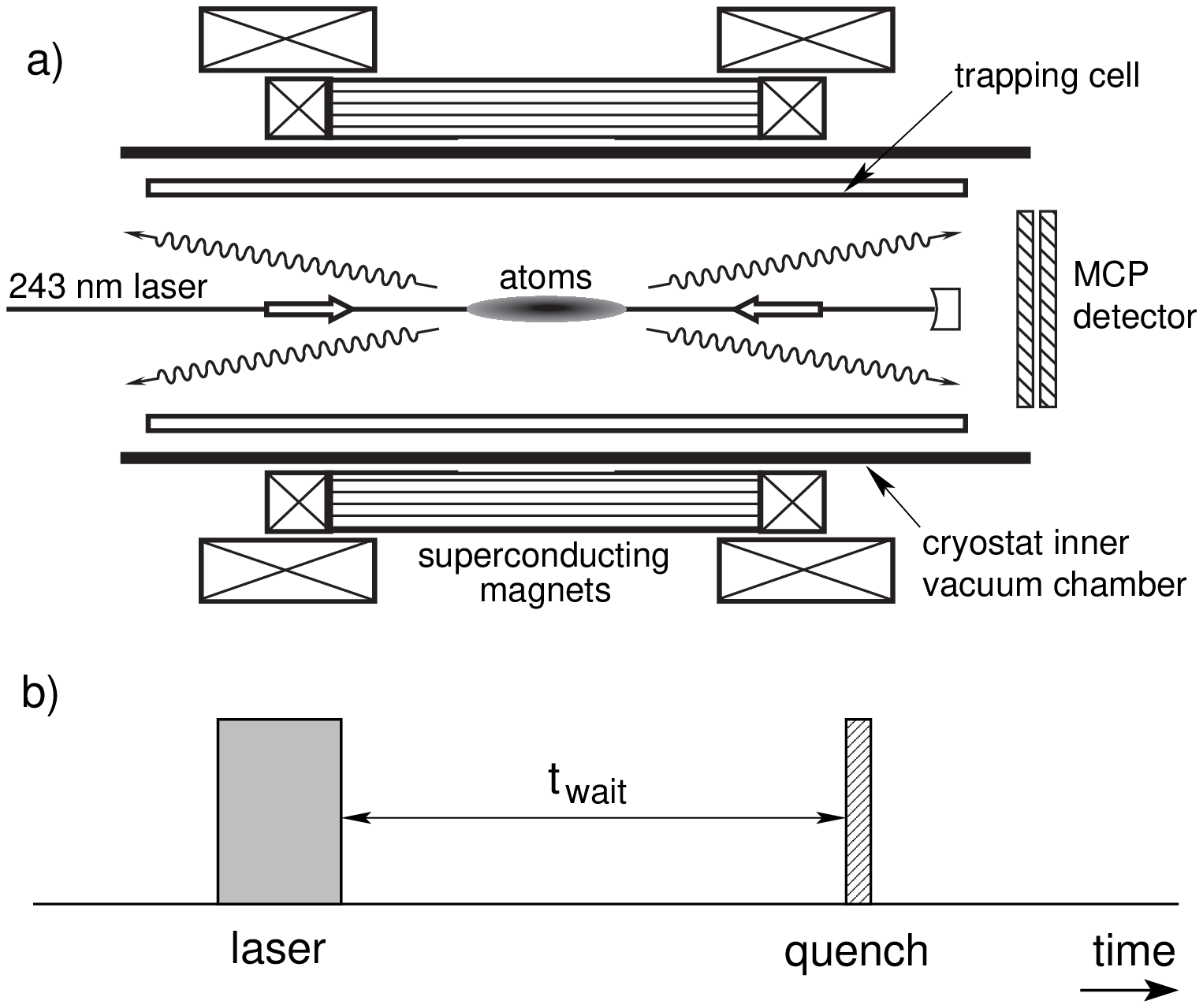,width=5in}
\vspace{.2in}
\caption{(a) A schematic of the cryogenic trapping apparatus.
Superconducting magnets produce the Ioffe-Pritchard trapping field.
The UV excitation laser is reflected back on itself to permit
Doppler-free two-photon excitation.  The apparatus is roughly
cylindrically symmetric about the trap axis, to which the laser is
aligned.  (b) Timing diagram for $1S$-$2S$ excitation spectroscopy.
After a $\sim2$~ms excitation pulse and an additional time $t_{\rm
wait}$, the metastables are Stark-quenched by an electric field pulse,
resulting in a burst of Lyman-$\alpha$ photons.  In a typical decay
measurement, $t_{\rm wait}$ is cycled through values between 1 and 92
ms.}
\label{apptim}
\end{figure}

The methods we use for trapping and cooling atomic H in a cryogenic
apparatus (Fig.~\ref{apptim}(a)) are described elsewhere
\cite{kfw98,gkf00,fkw98}.  Molecular hydrogen is dissociated in a
discharge chamber that is thermally anchored at one end to a dilution
refrigerator and opens to a trapping cell at the other end.  Atoms in
the $F=1,~m_F=1$ hyperfine state of the $1S$ manifold are captured in
a Ioffe-Pritchard magnetic trap \cite{pri83}, initially 500~mK deep.
With typical aspect ratios between 100:1 and 400:1, the cloud is
shaped like a highly elongated ellipsoid, cylindrically symmetric
about the trap axis.  We refer to the direction along the trap
axis as ``axial'' and to the perpendicular direction as ``radial.''  

If an atom has sufficient energy, it can escape over a magnetic field
saddlepoint located at one end of the trap.  The sample can be cooled
to a temperature of $\sim200~\mu$K by progressively lowering the
saddlepoint field, thus forcing evaporation \cite{hes86}.  If further
cooling is desired, rf-induced evaporation is employed to reach
temperatures as low as $\sim20~\mu$K \cite{phm88}.  The number of
atoms in the trap ranges from more than $10^{14}$ after initial
loading to $\sim10^{10}$ at the lowest temperatures.  For the
experiments described in this paper, the samples were not Bose
degenerate, and typical peak $1S$ densities ranged from
$10^{13}$~cm$^{-3}$ to $10^{14}$~cm$^{-3}$.


To generate metastable hydrogen, we excite the two-photon $1S$-$2S$
transition with a 243~nm laser \cite{cfk96}.  To achieve large
excitation rates, typically $\sim15$~mW of laser power is focused to a
radius of $\sim20$~$\mu$m near the trap minimum.  In accordance with
the two-photon selection rules, the metastables produced are in the
$F=1,~m_F=1$ state of the $2S$ manifold; except for a small
relativistic correction $(\sim10^{-4})$, these experience the same
magnetic trapping potential as the $1S$ atoms.  In the presence of a weak
electric field of strength $E$, the $2S$ state quenches at a rate
$\gamma_s=2800E^2~($cm$^2/$V$^2)$s$^{-1}$ due to Stark-mixing with the
$2P$ state\cite{bs77}.  To detect the metastables, a field of 10 V/cm
is applied, quenching the $2S$ atoms in a few microseconds.  A small
fraction of the resulting 122~nm Lyman-$\alpha$ photons are counted on
a microchannel plate (MCP) detector.  The metastable decay behavior is
observed by cycling through a series of different wait times between
excitation and quench pulses (Fig.~\ref{apptim}(b)).  During the trap
lifetime of the ground state sample, typically several minutes, we can
make many $2S$ decay measurements.  Samples of differing initial $2S$ density are obtained by stepping the laser frequency across the resonance.

For the measurements discussed here, the detection efficiency was
approximately $2\times10^{-6}$.  This was calibrated using a
230~$\mu$K sample where the initial number of $1S$ atoms was known
from bolometric density measurements \cite{kfw98} and the trap
geometry.  The sample was depleted by resonant $1S$-$2S$ excitation
and subsequent quenching to an untrapped state.  With appropriate
corrections, the detection efficiency can be inferred from the total
number of signal photons counted and the initial number of atoms
\cite{landhuis_thesis}.  The corrections account for other loss
mechanisms, for the branching ratio back to the trapped ground state,
and for reabsorption (``radiation trapping'') of the Lyman-$\alpha$
photons by the ground state cloud.  The photons emitted by the
quenched metastables have a frequency which is detuned from a
100~MHz-wide $1S$-$2P$ absorption resonance by only 1.1~GHz, and at
the densities achieved in our experiments, the effects of radiation
trapping are significant.  Using Monte Carlo simulations of
Lyman-$\alpha$ propagation in experimentally realistic cloud
geometries, it was determined that typically $\sim30$\% of potential
signal photons are absorbed and scattered in the atom cloud
\cite{landhuis_thesis}.  Other factors, such as the small detection
solid angle ($1.1\times10^{-2}$~sr), the transmission of two MgF$_2$
windows between the atoms and the MCP, and the MCP quantum efficiency,
lead to the small overall detection efficiency.

Based on the detection efficiency, we can estimate $N_{2S}$, the
population of the metastable cloud, and $n_{2S}$, the peak metastable
density in the trap.  In a very cold sample (87~$\mu$K), as many as
150 Lyman-$\alpha$ photons were observed on the MCP after a single
laser shot, implying $N_{2S}\simeq8\times10^7$.  The effective volume,
$V_{2S}=N_{2S}/n_{2S}$, of the metastable clouds for these samples is
$\sim1\times10^{-3}$~cm$^3$, which means that densities of
$\sim10^{11}$~cm$^{-3}$ have been achieved.


Due to scatter from optics inside the trapping cell, each laser pulse
induces some background fluorescence which is detected with low
efficiency at the MCP.  The background, which is typically small
compared to the metastable signal, decays in a few milliseconds to a
negligible level.  To establish a background correction for
our decay data, the laser is tuned far off resonance at the end of
each trapping sequence.  The background fluorescence is then recorded
for the same cycle of wait times used when generating metastables.

\begin{figure}
\vspace{0.5in}
\centering\epsfig{file=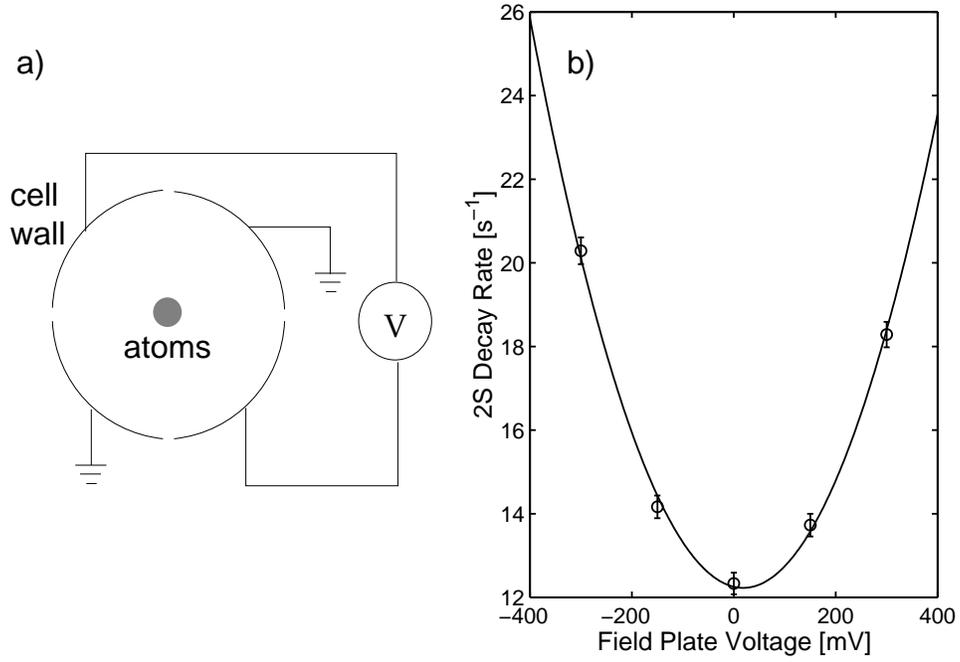,width=5in}
\vspace{.2in}

\caption{(a) Schematic of trapping cell in cross section, showing
electrical connections for the copper film electrodes used for both
quenching and stray field compensation.  The diameter of the trapping
space is 4~cm.  In the current design, compensation is possible for
only one spatial direction.  (b) Example of decay rate measurements
for different applied dc fields for a single trapped sample.  A
parabola is fit to the data points, indicating the quadratic
dependence of Stark quenching rate on the applied electric field.}

\label{efield}
\end{figure}

In earlier experiments, performed in an entirely nonmetallic cell, we
found that stray electric fields limited metastable lifetimes to 1~ms
or less.  In the current version of the apparatus, a copper film has
been added to the inside surface of the trapping cell.  The film is
thin enough to allow rf to penetrate the trapping cell from outside
coils, yet thick enough to significantly suppress stray dc fields.  As
depicted in Fig.~\ref{efield}(a), the copper film has been divided
into several contiguous electrodes which are used to apply both the
quench pulses and a compensating dc field.  The dc voltage which best
compensates for stray fields is found by measuring the metastable
decay rate in a low density sample for several applied voltages.  As
shown in Fig.~\ref{efield}(b), we observe the expected quadratic
dependence of the decay rate on electric field.  With optimal stray
field compensation, metastable lifetimes as long as 100~ms have been
observed, approaching the natural lifetime of the $2S$ state.

Our decay measurements have focused on metastable clouds generated in
four magnetic trap configurations, spanning a range of temperatures
and ground state densities.  These trap configurations are designated
Traps W, X, Y, and Z, and each has its own standard evaporation
sequence for sample preparation.  To increase the signal-to-noise
ratio, decay measurements were repeated in each trap over several
consecutive trap cycles.

\begin{figure}
\vspace{0.5in}
\centering\epsfig{file=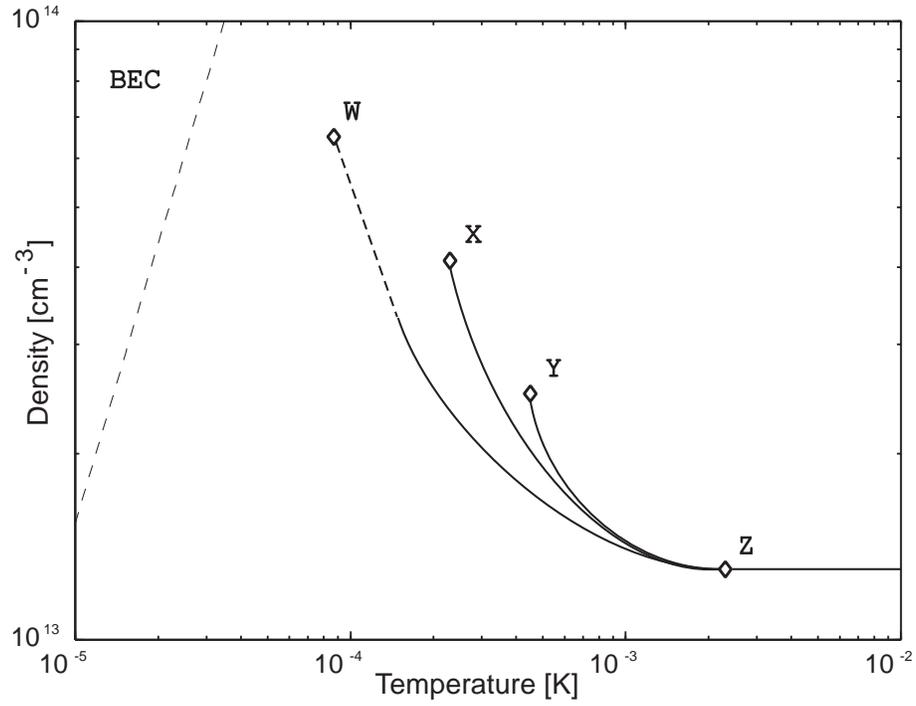,width=5in}
\vspace{.2in}

\caption{Phase space plot of H samples used in metastable decay
studies.  ``Density'' is the peak initial ground state density
prepared in the trap, and ``Temperature'' is the temperature of both
ground state and metastable clouds.  The approximate trajectory of the
samples during magnetic saddlepoint evaporation is indicated by solid
lines; for Trap W, a dashed line indicates the rf evaporation
trajectory.  A longer dashed line marks the boundary between the
Bose-condensed (BEC) and normal phases of the gas.}

\label{phasespace}
\end{figure}

Densities and temperatures of the standard samples and the approximate
trajectories by which they were created are depicted in
Fig.~\ref{phasespace} \cite{landhuis_thesis}.  To ensure that the
colder samples (W, X, and Y) had reached thermal equilibrium, they
were held undisturbed for 30~s after the end of the forced
evaporation.  The sample densities were measured to 20\% accuracy by
the bolometric method \cite{kfw98}, and the temperatures were
determined by a combination of methods.  For the coldest sample (W),
the temperature was calibrated to within 5\% by measuring the width of
the Doppler-sensitive $1S$-$2S$ line \cite{fkw98}.  The temperature of
Trap X was determined by comparing the widths of the Doppler-free
$1S$-$2S$ resonances of Traps W and X in the low density limit
\cite{cfk96}.  For Traps Y and Z, the temperature was established
using results from a numerical simulation of evaporative cooling
\cite{doy91}.  The temperature uncertainty for Traps X, Y, and Z is
estimated to be 10\%.

\boldmath
\section{Evidence for $2S$-$2S$ Two-Body Loss}
\unboldmath

\begin{figure}
\vspace{0.5in}
\centering\epsfig{file=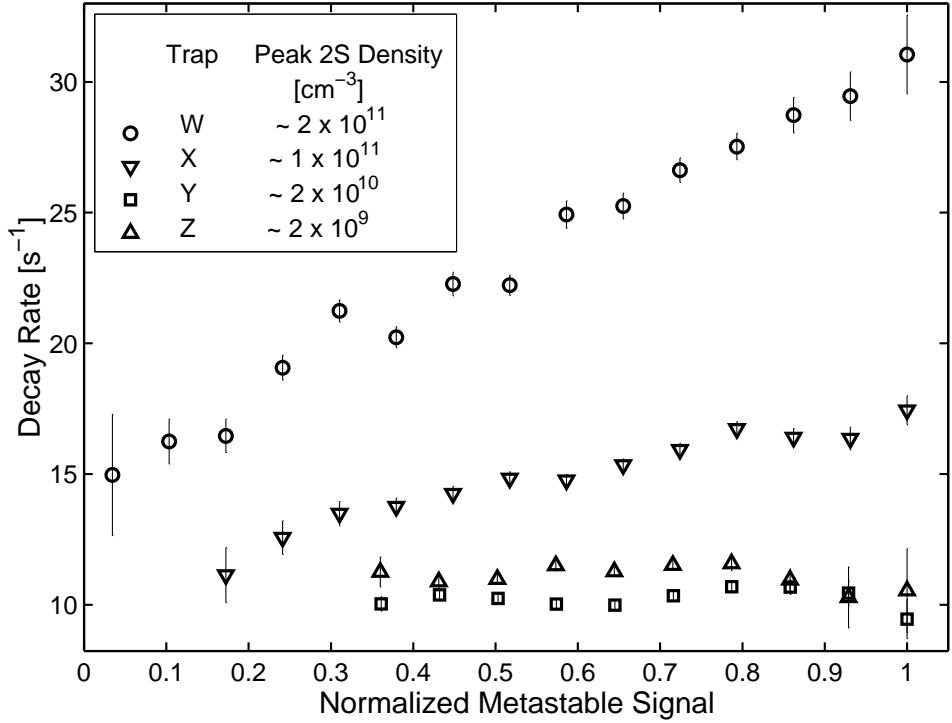,width=5in}
\vspace{.2in}

\caption{Decay rates determined by exponential fits to decay curves
recorded at different metastable signal levels in Traps W, X, Y,
and Z.  The bottom edge of the plot corresponds to the
natural $2S$ decay rate of 8.2~s$^{-1}$, and the error bars are
derived assuming only statistical uncertainty in the decay data.
Different signal levels were achieved by varying the laser detuning
and allowing the ground state sample to decay.  The metastable signal,
which is approximately proportional to the metastable density, has
been normalized to the peak signal observed in each trap; estimates of
the peak $2S$ density corresponding to the peak signal are given in
the legend.  The metastable clouds produced in Traps W and X show a
clear density dependence of the decay rate.}

\label{decayratesWXYZ}
\end{figure}

Our investigation of inelastic $2S$-$2S$ collisions was initiated by
observations of how the decay behavior depended on metastable density.
In each of the standard traps the $2S$ density was varied by
generating metastables at different laser detunings and also by
exciting different $1S$ densities as the ground state sample decayed.
The decay curves were binned according to the total number of
metastable signal counts in each decay curve.  Since, to a first
approximation, the signal is proportional to the metastable density,
this means that decay curves corresponding to approximately the same
initial $2S$ density were averaged together.  For a preliminary
analysis, the average decay curves were fit to a simple exponential
function, with only the decay rate and initial amplitude as free
parameters.  Figure~\ref{decayratesWXYZ} shows the results of this
analysis.  In the two coldest traps, W and X, the decay rate clearly
increases with increasing metastable density, consistent with the
presence of two-body inelastic loss processes.  Furthermore, among the
four traps, the decay rate is largest in Trap W, where the highest
metastable densities are achieved.

The peak $2S$ densities achieved in each trap can be estimated from
the number of signal counts observed, but the uncertainty is large,
reflecting significant uncertainties in both the metastable cloud
shape and the detection efficiency.  The shape of the $2S$ cloud
depends sensitively on the location of the laser beam waists with
respect to the magnetic trap.  For the density values listed in
Fig.~\ref{decayratesWXYZ}, a detection efficiency of $2\times10^{-6}$
was assumed, and the counterpropagating laser beams were assumed to be
perfectly overlapped and well-aligned with the atom cloud; with a
plausible misalignment, the actual densities could be a factor of 3
smaller.  

\begin{figure}
\vspace{0.5in}
\centering\epsfig{file=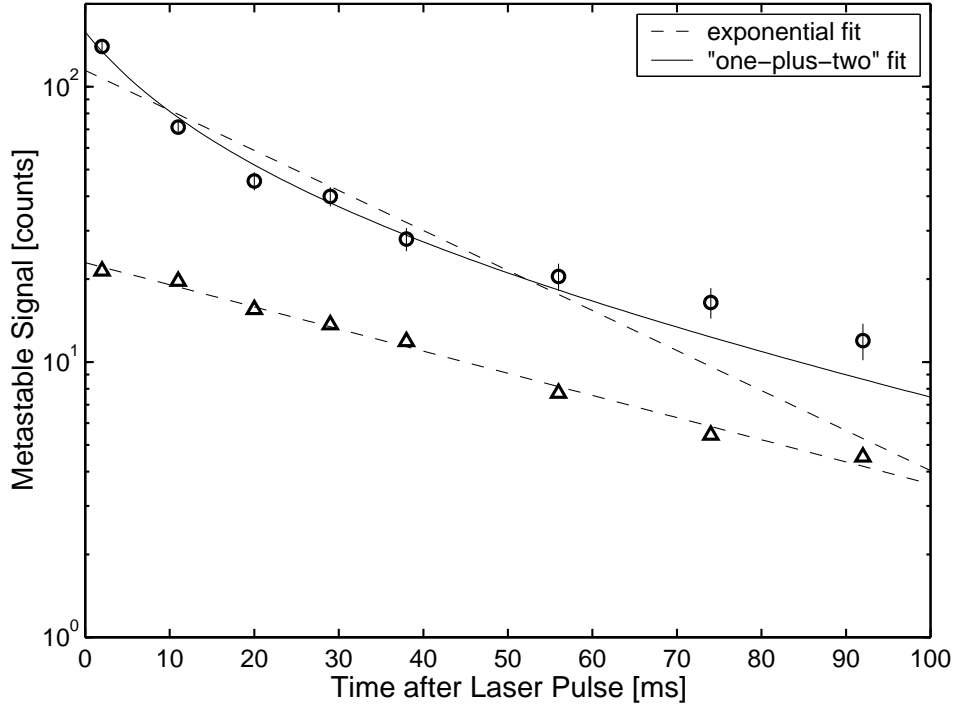,width=5in}
\vspace{.2in}

\caption{Average decay curves for an initial $2S$ density near the
maximum achieved in Trap W (circles) and for an initial density in
Trap W more than six times smaller (triangles).  An exponential decay
describes the low density data well, but the the high density data
deviates significantly from the best fitting exponential decay.  A
simple ``one-plus-two'' model which has a fixed one-body decay parameter
and a free two-body decay parameter fits the first $\sim50$~ms of the
decay curve well.  The model assumes a static cloud shape; in reality,
the $2S$ cloud changes shape as it decays, leading to deviation from
the model at long wait times.}

\label{decaycurves}
\end{figure}

As illustrated in Fig.~\ref{decaycurves}, a comparison of the average
decay curves at high and low metastable signal provides further
evidence for two-body loss at high densities.  At low $2S$ densities,
the decay curves are well described by an exponential decay.  By
contrast, the decay curves at high $2S$ densities deviate
significantly from a simple exponential.  They are better fit by a
``one-plus-two'' model with a free two-body decay parameter and a
one-body decay rate fixed at its low density value.  The one-plus-two
model is described further in Sec.~\ref{sec:determination}.

In principle, three-body inelastic collisions could also contribute to
the observed density dependence of the decay.  However, the three-body
loss rate constant for metastables would have to be unexpectedly large
to play a significant role at our experimental densities
($\alt10^{11}$~cm$^{-3}$) \cite{for01}.  In the following analysis,
we assume that two-body effects dominate.

\section{Determination of the Two-Body Loss Rate Constant}
\label{sec:determination}


In this section, we derive the rate constant $K_2$ for two-body loss
of metastables based on decay data of the form $s(t)$, the number of
signal counts observed when quenching at wait time $t$.  The constant
is defined as follows to give the rate of change of the local $2S$
density $n_{2S}({\bf r})$ due to two-body processes:
\begin{equation}
\label{eq:K2definition}
\Bigl. \dot{n}_{2S}({\bf r}) \Bigr|_{\rm two-body} = - K_{2} n_{2S}^2({\bf r}).
\end{equation}
If the relative distribution of metastables in the trap is not
changing in time (``static approximation''), $K_2$ is the product of
the metastable detection efficiency, a factor describing the cloud
geometry, and a two-body decay parameter extracted from fits to
$s(t)$.  After analyzing the decay process in the static
approximation, we introduce dynamic corrections to obtain final
results for $K_2$.

Given the geometry of our apparatus and the ground state densities
achieved, it is a reasonable first approximation to assume that the
metastable cloud does not change shape as it decays.  The initial
extent of the metastable cloud along the trap axis, about 2~cm between
the $1/e$ points, is defined by the depth of focus of the laser.  This
is shorter than the thermal length of the ground state cloud, which
ranges from 4~cm in Trap W to 23~cm in Trap Z.  On the time scale of
the decay, the $2S$ atoms collide frequently with $1S$ atoms.  At the
peak ground state density in Trap W, for example, the $2S$ mean free
path is 100~$\mu$m, and the mean time between collisions is 70~$\mu$s.
(To calculate these values, we use the theoretical value for
$a_{1S-2S}$, the elastic $1S$-$2S$ scattering length \cite{osw99}.)
Thus, the metastables diffuse slowly along the trap axis.  Numerical
simulations of the time evolution of the $2S$ cloud have shown that
over a cloud lifetime of 100~ms, the shape of the initial axial
distribution does not change dramatically \cite{landhuis_thesis}.  In
the radial direction, the cloud shape evolves even
less, although for a different reason.  Due to the geometry of the
magnetic trap, the metastable cloud is cylindrically symmetric and
confined to the same thermal radius as the ground state cloud.  This
thermal radius, about 110~$\mu$m for Trap W, is comparable to the $2S$
mean free path.  In a time shorter than the 2~ms excitation pulse, the
metastables establish an equilibrium spatial distribution across the
short radial dimension.  The relative spatial distribution of
metastables is quasi-static, then, because (1) axial diffusion through
the $1S$ gas is slow and (2) radial equilibrium exists over the
lifetime of the $2S$ cloud.

In the static approximation, the rate of signal decay can be described
by the differential equation
\begin{equation}
\dot{s}=-\alpha_1 s - \alpha_2 s^2,
\label{eq:oneplustwodiffeq}
\end{equation}
where $\alpha_1$ and $\alpha_2$ are, respectively, parameters
corresponding to one- and two-body decay.  The solution to
Eq.~\ref{eq:oneplustwodiffeq} is
\begin{equation}
s(t) = \frac{\alpha_1 A_o e^{-\alpha_1 t}}
{\alpha_1 + \alpha_2 A_o (1 - e^{-\alpha_1 t})},
\label{eq:oneplustwo}
\end{equation} 
which we have named the ``one-plus-two'' model.  The parameter
$\alpha_1$ is the one-body decay rate, the total rate of decay due to
processes which are independent of $2S$ density.  Contributions to the
one-body rate include the natural decay of the $2S$ state,
quenching due to stray electric fields, and inelastic collisions with
the background $1S$ gas.  

\begin{figure}
\vspace{0.5in}
\centering\epsfig{file=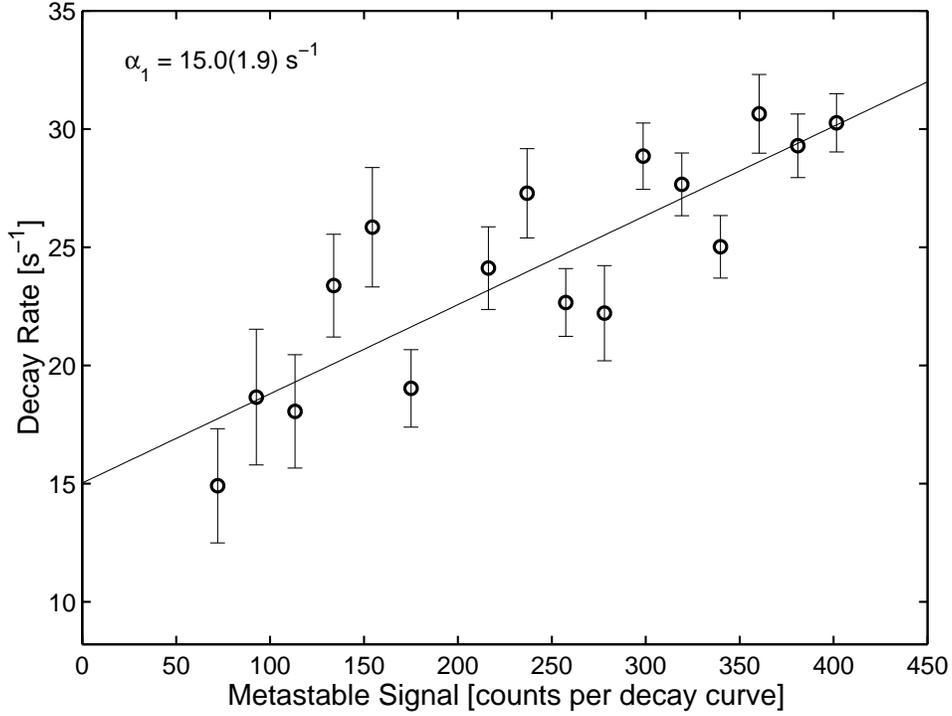,width=5in}
\vspace{.2in}

\caption{Determination of $\alpha_1$ by linear extrapolation of
exponential decay rates to zero metastable signal.  The decay data in
this example was recorded in Trap W, and the bottom edge of the plot
represents the natural decay rate.  To avoid possible systematic
errors due to a changing ground state density, the data set includes
only those decay curves recorded in the first 9.6~s following
excitation.  The error bars on the decay rates assume only statistical
errors in the decay curves; the uncertainty in $\alpha_1$ is based on the
scatter of decay rate values with respect to the linear fit.}

\label{alpha1determination}
\end{figure}

For a decay measurement with a very high signal-to-noise ratio, one
could extract $\alpha_1$, $\alpha_2$, and $A_o$ simultaneously by a
fit to the ``one-plus-two'' model.  The parameters $\alpha_1$ and
$\alpha_2$ are highly correlated, however.  With the modest
signal-to-noise ratio in our data, a consistent convergence of the fit
is not assured.  Thus, we employ an alternate method of analysis.  For
a given trap and $1S$ density, the value for $\alpha_1$ is determined
by extrapolating the best-fitting exponential decay rates to zero
metastable density (Fig. \ref{alpha1determination}).  This leaves two
free parameters in the one-plus-two model: $A_o$, the signal which
would be observed at $t=0$, and $\alpha_2$, from which $K_2$ can be
calculated.

By integrating Eq.~\ref{eq:K2definition} over the entire trap, we
obtain an expression for the total number of metastables lost per unit
time due to two-body processes:
\begin{equation}
\Bigl. \dot{N}_{2S} \Bigr|_{\rm two-body} = - K_{2} n_{o,2S}^2 \int
f_{2S}^2({\bf r})\,d^3{\bf r}.
\label{eq:entiretraploss}
\end{equation}
Here, $N_{2S}$ is the number of $2S$ atoms in the trap, and 
$f_{2S}({\bf r}) = n_{2S}({\bf r})/n_{o,2S}$ is the normalized spatial
distribution function for metastables, where $n_{o,2S}$ is the peak
$2S$ density.  Since $s=\epsilon N_{2S}$, where
$\epsilon$ is the detection efficiency, and since
\begin{equation}
\Bigl. \dot{s} \Bigr|_{\rm two-body} = - \alpha_2 s^2,
\end{equation}
it follows that
\begin{equation}
K_2 = \zeta \epsilon \alpha_2
\label{eq:staticK2}
\end{equation}
in the static approximation, where $\zeta$ is a geometry factor defined by
\begin{equation}
\zeta=\frac{\int f_{2S}^2({\bf r})\,d^3{\bf r}}{\left[ \int
f_{2S}({\bf r})\,d^3{\bf r}\right]^2}.
\label{eq:zeta}
\end{equation}
To obtain Eq.~\ref{eq:staticK2}, we have used the fact that
$n_{o,2S}=N_{2S}/ \int f_{2S}({\bf r})\,d^3{\bf r}$.  The quantity
$\zeta$, which has units of volume, is a measure of the spatial extent
of the metastable cloud; $\zeta$ decreases as $f_{2S}({\bf r})$
becomes more sharply peaked.


Thus to determine $K_2$ from our decay data it is necessary to know
$\epsilon$ and the shape of the metastable cloud.  Our method for
determining $\epsilon$ was described in Sec.~\ref{sec:methods}.  To
calculate $\zeta$ for various trap configurations and temperatures, we
employed a Monte Carlo simulation of $1S$-$2S$ excitation in our trap
\cite{wil02}.  The simulation calculates the excitation
lineshape associated with atoms in a volume slice perpendicular to the
trap axis.  In addition to the position of the volume slice along the
axis, the inputs to the simulation include the laser field geometry,
the laser pulse length, the trap shape, the sample temperature, and
the peak $1S$ density.  The lineshape is calculated by choosing atoms
randomly from a thermal distribution, computing their trajectories
during the laser pulse, and then finding the contribution of each atom
to the spectrum by Fourier-transforming the time-varying laser field
amplitude seen by the atom.  The effects of the $1S$-$2S$ cold
collision shift \cite{kfw98} are incorporated by allowing the
resonance frequency to vary over the trajectory in accordance with the
local $1S$ density.  After repeating the lineshape calculation for
many slices covering the length of the $2S$ cloud along the trap axis,
the axial distribution of metastables at $t=0$ can be extracted for
any given laser detuning.  Under the assumption that the metastables
establish an equilibrium radial distribution during the pulse,
$f_{2S}({\bf r})$ for the entire trap is easily calculated from the
axial distribution.

An additional numerical calculation corrects the lineshape simulation
results to account for photoionization, the promotion of $2S$
electrons to the continuum by the excitation laser.  In a cold, dense
sample, such as those prepared in Trap W, the metastables spend a
relatively long time in the high-intensity UV laser field, and the
fraction of metastables lost to photoionization can exceed 30\%
\cite{landhuis_thesis}.

Because the lineshape simulation has given results in good agreement
with experimental $1S$-$2S$ spectra for a wide range of trap shapes,
densities, and temperatures, this method of calculating $f_{2S}({\bf
r})$ is believed reliable.  For each spatial distribution function,
$\zeta$ is computed by numerical evaluation of Eq.~\ref{eq:zeta}.
Values for $\zeta$ range from approximately $2\times10^{-3}$~cm$^3$ in
Trap W to 0.2~cm$^3$ in Trap Z.  Depending on the laser detuning, an
imperfect overlap of the counterpropagating laser beam focuses along
the trap axis may cause $\zeta$ to be 50-100\% larger.  Uncertainty in
the knowledge of the laser field geometry is the principal source of
uncertainty in $\zeta$.


The experimental results for $K_2$, calculated on the basis of
Eq.~\ref{eq:staticK2}, are listed in the ``static approximation''
column of Table~\ref{tab:K2results}.  For each standard trap, the
accumulated decay curves were binned together according to laser
detuning, and the resulting average curves were fit with the
one-plus-two model to obtain $\alpha_2$ as a function of detuning.
Next, the values for $\alpha_2$ were multiplied by the detection
efficiency and the value of $\zeta$ appropriate for the detuning.  The
resulting values for $K_2$ were found to be largely independent of
laser detuning, indicating that variations in metastable cloud shape
with detuning are reasonably accounted for by our lineshape
simulation.  For Traps W and X, the weighted averages across all
detunings are reported in Table~\ref{tab:K2results}.  In the warmer
samples, where the $2S$ density is too low to observe two-body
collisions, the weighted average for $K_2$ is consistent with zero.
For these, Table~\ref{tab:K2results} reports upper limits for $K_2$
based on linear addition of the uncertainties in $\alpha_2$,
$\epsilon$, and $\zeta$.

The error bars for Traps W and X are primarily due to systematic
uncertainties.  These errors are asymmetric due to asymmetric
uncertainty in the detection efficiency and the fact that laser
misalignments generally increase $\zeta$.  The most important
contributions to the overall error in $K_2$ come from the detection
efficiency and uncertainty in the axial overlap of the laser focuses.
In Trap W, for example, $\epsilon=1.8\times10^{-6}$ with approximate
relative uncertainties of 30\% on the lower side and 60\% on the upper
side.  The focus overlap uncertainty contributes 75\% to the upper
relative error.  Other sources of uncertainty include the error in the
determination of $\alpha_2$~$(\pm12\%)$ and contributions of a few
percent each due to uncertainties in the beam radius, cold collision
shift parameter, offset of the trap minimum and laser focuses, ground
state density, and sample temperature.  With some minor exceptions,
the various uncertainties are assumed to be uncorrelated, and the
relative errors we report for $K_2$ are essentially quadrature sums of
the contributing errors.


The static approximation results need to be refined to account for
change in the metastable spatial distribution while the cloud decays.
There is a flattening of the initial axial distribution due to
diffusion and to two-body loss, which occurs preferentially where the
$2S$ density is highest.  This flattening of the distribution in time
causes the initial decay of the metastable population to be somewhat
less steep than it would be if the cloud shape were completely static.
Thus, the value of $\alpha_2$ determined by fitting a decay curve to
the one-plus-two model is smaller than the effective value of
$\alpha_2$ which accounts for dynamics when substituted into
Eq.~\ref{eq:staticK2}.

To calculate corrections to the static approximation analysis, a
numerical simulation of the decaying metastable cloud was developed.
For each standard sample, the simulation was repeated many times
assuming different laser detunings.  The starting point for each
simulation run was a cloud with distribution $f_{2S}({\bf r})$,
calculated by the method described above, and an initial total number
of metastables $N_{2S}(0)=A_o/\epsilon$, where $A_o=s(0)$ was
determined from experimental data.  The simulation then evolved both
dynamic and static clouds forward in time from the initial condition.
For the dynamic cloud, the axial distribution was adjusted locally at
each time step for losses due to one-body processes (parametrized by
$\alpha_1$), two-body processes (parametrized by a putative value for
$K_2$), and for diffusion of $2S$ atoms in the $1S$ background.  For
the static cloud, $N_{2S}$ was adjusted at each time step for one- and
two-body losses integrated over the whole cloud, but the spatial
distribution remained fixed.  After evolving the clouds over 100~ms,
the simulated static and dynamic metastable decay curves were fit by
the one-plus-two model.  The ratio $\alpha_{2,{\rm
static}}/\alpha_{2,{\rm dynamic}}$ from the simulation was used as a
correction factor for the experimentally determined $\alpha_2$ values.
New average values for $K_2$ were then calculated for each
trap.  It was found that the static approximation values of $K_2$ must
be multiplied by 1.35 and 1.30, respectively, to correct for dynamics
in Traps W and X.  By varying the inputs to the decay simulation, the
uncertainty in the overall correction factors was estimated to be
10\%.  For Traps Y and Z, a conservative upper limit for the
correction factor is 2.  The final results for $K_2$ are summarized in
the last column of Table~\ref{tab:K2results}.

\begin{figure}
\vspace{0.5in}
\centering\epsfig{file=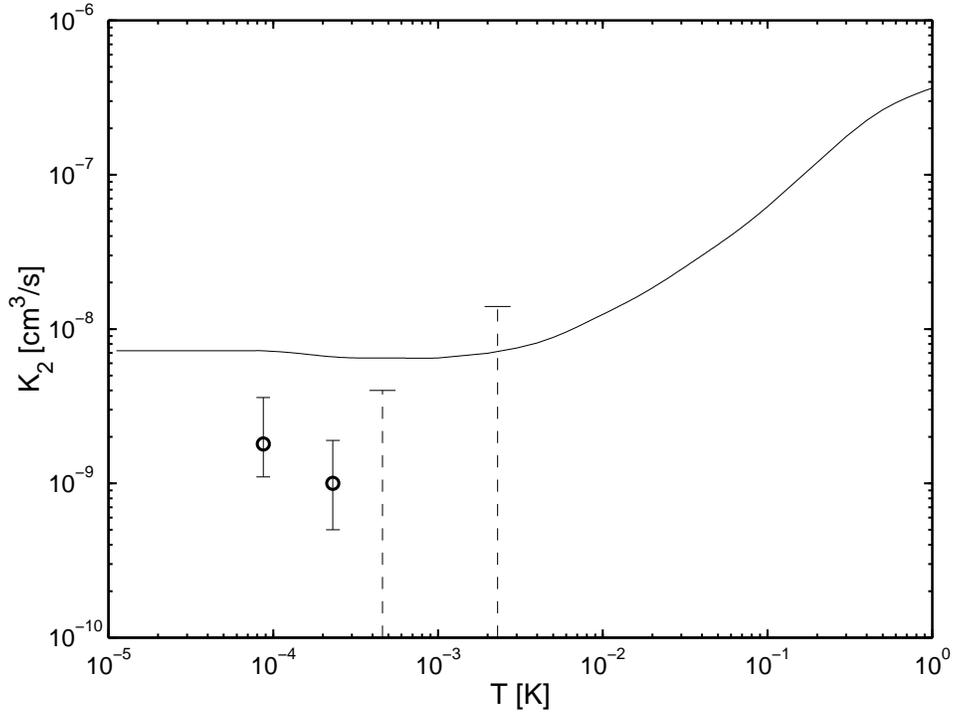,width=5in}
\vspace{.2in}

\caption{Comparison of experimental results (circles) for $K_2$ with
the theoretical calculations (solid line) of Forrey and collaborators.
At the temperatures where only upper limits were determined, dashed
lines indicate the possible range of $K_2$.  Since the uncertainties
in the experimental points are highly correlated, the experimental
results suggest that $K_2$ has a significant temperature dependence
below 230~$\mu$K.}

\label{K2theoryandexp}
\end{figure}

Theoretical $2S$-$2S$ collision rates at low temperatures were
calculated recently by Forrey and collaborators \cite{fjs02},
extending their previous work for high temperatures \cite{fcd00}.  The
calculations indicate that the most prevalent inelastic process in a
cold metastable cloud is double excitation transfer, ${\rm H}(2S) +
{\rm H}(2S) \longrightarrow {\rm H}(2P) + {\rm H}(2P)$.  Next in
importance are the two possible ionization processes: associative
ionization, in which the molecular ion ${\rm H}_{2}^+$ is formed, and
Penning ionization, where the internal energy of one metastable atom
causes the ionization of a second metastable.  Since processes
involving $2S$ hyperfine state changes are mediated by comparatively
weak magnetic dipole interactions, they are expected to be negligible
in relation to excitation transfer and ionization.  Thus, a
theoretical estimate for $K_2$ is obtained by summing the loss rate
constants for double excitation transfer and ionization.  For each
process, the loss rate constant equals $2\langle \sigma v \rangle$,
where $\sigma$ is the cross section for the process, $v$ is the
relative velocity of two atoms, the angle brackets indicate a thermal
average, and the factor of 2 accounts for the fact that two
metastables are lost in each collision.  The theoretical curve for
$K_2$ as a function of temperature is shown alongside the experimental
values in Fig.~\ref{K2theoryandexp}.

The uncertainty in the theoretical results is difficult to estimate.
In its current state of development, the theory for $2S$-$2S$
collisions neglects hyperfine structure and assumes zero coupling
between the relative orbital angular momentum of the nuclei and the
angular momentum of the electrons.  It is not yet certain whether this
assumption is justified for spin-polarized metastables at low
temperatures \cite{for01}.  Furthermore, the theory calculations
assume zero magnetic field, while in the experiment the trapped
metastables experience a field of a few gauss.  Nevertheless, from
Fig.~\ref{K2theoryandexp} we can conclude that the present theory
correctly predicts $K_2$ at low temperatures to within an order of
magnitude.

An intriguing aspect of the experimental results for $K_2$ is the
apparent temperature dependence between 87~$\mu$K and 230~$\mu$K.
Although the absolute uncertainty is large for each temperature point,
we believe that the temperature dependence is real.  The error bars
are primarily due to systematic uncertainties which affect both points
in approximately the same way.  When the correlated errors are
removed, the ratio of $K_2$ at 87~$\mu$K to $K_2$ at 230~$\mu$K is
found to be $1.9\pm0.4$.  This result contrasts with the theoretical
prediction that $K_2$ is virtually independent of temperature in this
regime, consistent with Wigner threshold behavior.  Further
theoretical and experimental work will be necessary to understand the
temperature dependence of inelastic $2S$-$2S$ collisions near $T=0$.


\boldmath
\section{Upper Limit for the $1S$-$2S$ Loss Rate Constant}
\unboldmath 


In addition to the two-body loss channels discussed above,
metastables are lost from the trap by several one-body mechanisms.
These include the natural decay of the $2S$ state by spontaneous
emission of two photons, quenching due to stray electric fields, and
loss due to inelastic $1S$-$2S$ collisions.  In a collision with a
ground state atom, a metastable may be lost by a transition to either
an untrapped $2S$ state (hyperfine-changing collision) or to a $2P$
state (excitation transfer).  If inelastic $1S$-$2S$ collisions are
sufficiently probable, they should result in a dependence of the
observed one-body decay parameter $\alpha_1$ on the peak ground state
density $n_{1S,o}$.

\begin{figure}
\vspace{0.5in}
\centering\epsfig{file=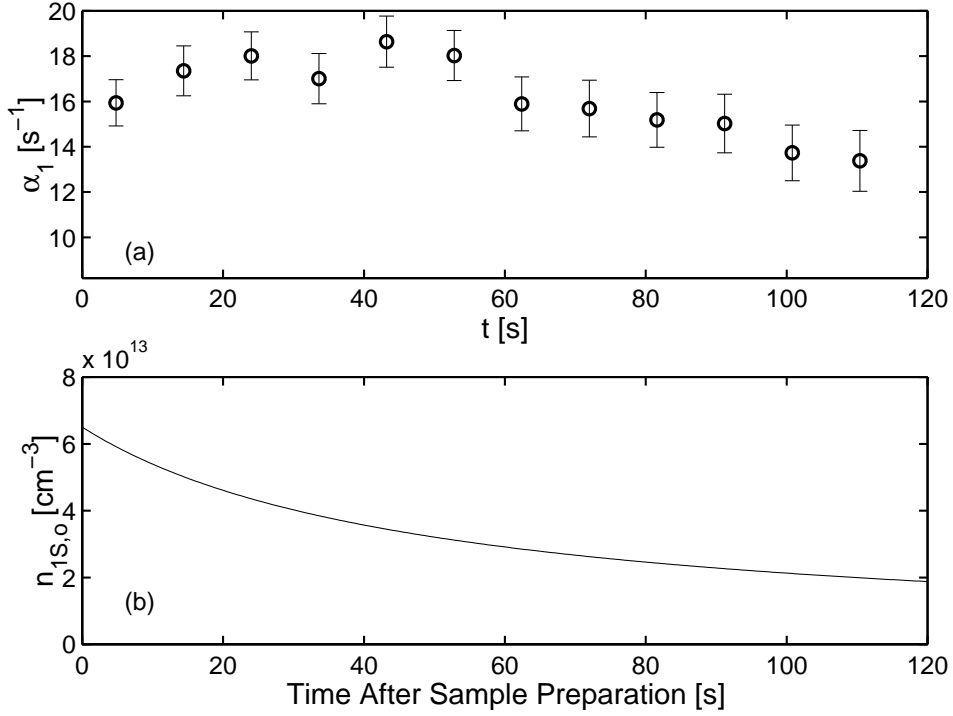,width=5in}
\vspace{.2in}

\caption{(a) Weighted average of the one-body metastable loss rate
$\alpha_1$ as a function of time after preparation of the ground state
sample in Trap~W.  The loss rate $\alpha_1$ has been determined by
extrapolation of exponential decay rates to zero metastable signal
(see Fig.~\ref{alpha1determination}).  The bottom edge of the plot
corresponds to the natural decay rate, and error bars include only
statistical uncertainties.  (b) Peak $1S$ density calculated as a
function of time in Trap W assuming that the only significant loss
mechanism is $1S$-$1S$ dipolar decay.  Comparison of (a) and (b)
suggests that some of the one-body loss may be due to inelastic $1S$-$2S$
collisions.}

\label{alpha1vstW}
\end{figure}

To search for evidence of inelastic $1S$-$2S$ collisions, the decay
curves taken in the standard trap configurations were organized into
consecutive time bins of 9.6~s duration.  Next, a value for
$\alpha_1$ was determined for each bin by extrapolating the decay
rate to zero metastable signal, as in Fig.~\ref{alpha1determination}.
Then, for each trap, $\alpha_1$ was plotted as a function of time over
the 120~s period following initial preparation of the sample (see
Fig.~\ref{alpha1vstW}(a), for example).  During this time, the ground
state density decays significantly due to $1S$-$1S$ dipolar decay
(Fig.~\ref{alpha1vstW}(b)).

As shown in Fig.~\ref{alpha1vstW}, the data suggests that some component
of one-body loss decreases as the ground state density decreases.
However, since the trend is not large relative to uncertainties, it is
inconclusive whether significant loss due to $1S$-$2S$ collisions
exists in our metastable clouds.  We have not attempted to fit the
time (or density) dependence of $\alpha_1$ to a model but instead have
determined conservative upper limits for the metastable loss rate due
to collisions with the ground state background.


An upper limit on the total inelastic contribution $\alpha_{1,{\rm inel}}$
to the one-body decay rate is found by subtracting the natural decay
rate $(\gamma_{\rm nat}=8.2$~s$^{-1})$ from the total one-body decay rate:
\begin{equation}
\alpha_{1,{\rm inel}} < \alpha_1 - \gamma_{\rm nat}.
\label{eq:alphainellimit} 
\end{equation}
Analogous to Eq.~\ref{eq:K2definition}, we define a rate constant
$K_{12}$ for the metastable loss rate due to inelastic $1S$-$2S$
collisions, such that
\begin{equation}
\Bigl. \dot{n}_{2S}({\bf r}) \Bigr|_{1S-2S} = - K_{12}\,n_{1S}({\bf
r}) n_{2S}({\bf r}).
\label{eq:K12definition}
\end{equation}  
Using a static approximation for the metastable cloud,
Eq.~\ref{eq:K12definition} can be integrated over the entire trap to
find the relation between $K_{12}$ and $\alpha_{1,{\rm inel}}$:
\begin{equation}
K_{12} = \frac{\alpha_{1,{\rm inel}} V_{2S}}{n_{1S,o} Q_{12}},
\end{equation} 
where $V_{2S}=\int f_{2S}({\bf r})\,d^3{\bf r}$, and $Q_{12} = \int
f_{1S}({\bf r}) f_{2S}({\bf r})\,d^3{\bf r}$.  By combining an
experimental upper limit for the initial value of $\alpha_{1,{\rm
inel}}$ in each trap, a numerically computed upper limit for the ratio
$V_{2S}/Q_{12}$, and the measured peak density, we have obtained the
upper limits on $K_{12}$ listed in Table~\ref{tab:K12results}.


These upper limits are much larger than the theoretical rate constant
for $1S$-$1S$ dipolar decay, $g = 1.2\times10^{-15}$~cm$^3$/s
\cite{skv88}.  Since the total rate constant for $1S$-$2S$
hyperfine-changing collisions is expected to be comparable to $g$, our
results admit the possibility that $1S$-$2S$ excitation transfer
collisions are much more probable than hyperfine-changing collisions.
A first theoretical calculation of the relevant cross sections in this
temperature regime is currently in progress \cite{zyg02}.

\section{Conclusion}

In summary, we have completed the first determinations of loss rates
due to inelastic collisions in a cold, trapped metastable H gas
coexisting with a ground state background.  The experimental values
for the two-body loss rate $K_2$ agree to order of magnitude
$(\sim10^{-9}$~cm$^3$/s) with the present theory for $2S$-$2S$
collisions.  Further refinements to the theory will be necessary to
explain an apparent increase in $K_2$ with decreasing temperature
below 230~$\mu$K.  Our metastable decay measurements also show that
the total rate constant for inelastic $1S$-$2S$ collisions is not
larger than $\sim10^{-13}$~cm$^3$/s at the temperatures probed below
1~mK.

The present results pertain to total $2S$-$2S$ and $1S$-$2S$ inelastic
collision rates.  In future experiments with an enhanced
Lyman-$\alpha$ detection efficiency and more complete suppression of
stray electric fields, it may be possible to measure the rates due to
excitation transfer processes alone by directly observing the
fluorescence of $2P$ states generated in collisions.

With the ability to create clouds of $10^7$-$10^8$ trapped metastables
and a preliminary understanding of their inelastic collisions, the
prospects are bright for fruitful new spectroscopic experiments
involving the $2S$ state.  Recently, our group has performed
spectroscopy of the H $2S$-$4P$ transition in a magnetic trap for the
first time.  Absorption spectroscopy on this and other single-photon
transitions provide new tools for studying interactions of hydrogen
states at low temperatures.  It will also soon be possible to excite
narrow two-photon transitions from the $2S$ state, setting the stage
for precision frequency measurements which probe fundamental physics.

\section*{Acknowledgments}

We are grateful to Alexander Dalgarno, Robert C. Forrey, Dale
G. Fried, Piotr Froelich, and Thomas C. Killian for helpful conversations.  We
also thank Walter Joffrain for his assistance in the laboratory.  This
research was supported by the National Science Foundation and the
Office of Naval Research.


%
%
%


%
%
\begin{table}
\begin{center}
\begin{tabular}{ldcc}
& & Static approx. & With corrections \\ 
Trap & $T$~(mK) & $K_2$~$(10^{-9}$~cm$^3$/s) & $K_2$~$(10^{-9}$~cm$^3$/s) \\
\hline
& & & \\
W & 0.087 &$1.4^{+1.3}_{-0.5}$ & $1.8^{+1.8}_{-0.7}$ \\ 
& & & \\
X & 0.23 & $0.74^{+0.70}_{-0.35}$ & $1.0^{+0.9}_{-0.5}$\\
& & & \\
Y & 0.45 & $<2$ & $<4$ \\ 
& & & \\
Z & 2.3 & $<7$ & $<14$ \\ 
\end{tabular}
\end{center}
\caption{Results for $K_2$ in the static approximation and including
dynamic corrections.  Sources of uncertainty are discussed in the text.}
\label{tab:K2results}
\end{table}
\begin{table}
\begin{center}
\begin{tabular}{ldd}
Trap & $T$~(mK) & $K_{12}$~$(10^{-13}$~cm$^3$/s) \\ \hline
W & 0.087 &$<5$ \\
X & 0.23 & $<3$ \\
Y & 0.45 & $<2$ \\ 
Z & 2.3 & $<11$ \\ 
\end{tabular}
\end{center}
\caption{Experimental upper limits for $K_{12}$ in the static approximation.}
\label{tab:K12results}
\end{table}

\end{document}